\documentclass[conference]{IEEEtran}
\usepackage{cite}

\ifCLASSINFOpdf
   \usepackage[pdftex]{graphicx}
   \graphicspath{{./figures/}}
   \DeclareGraphicsExtensions{.pdf,.jpeg,.png}
\else

\fi
\usepackage{amsmath}
\usepackage{array}
\usepackage{fixltx2e}
\usepackage{color,soul}
\usepackage{subfig}
\usepackage{float}

\begin{document}
%
\title{A Programmable CMOS Transceiver for Structural Health Monitoring}

\author{\IEEEauthorblockN{Xinyao Tang, Haixiang Zhao, and Soumyajit Mandal}
\IEEEauthorblockA{Department of Electrical Engineering and Computer Science\\
Case Western Reserve University, Cleveland, OH 44106\\
Email: sxm833@case.edu}}


%


\maketitle

\begin{abstract}
We describe a highly-integrated CMOS transceiver for active structural health monitoring (SHM). The chip actuates piezoelectric transducers and also senses ultrasound waves received by the same or another transducer. The transmitter uses an integer-$N$ frequency synthesizer and pulse-width modulation (PWM) to generate low-distortion, band-limited waveforms up to 12.7~V$_{pp}$ with center frequency from $\sim$0.1-2.75~MHz. The integrated offset-canceling fully-differential receiver has programmable gain and bandwidth, and uses quadrature demodulation to extract both amplitude and phase of the received waveforms for further signal processing. The transceiver was fabricated in a 0.5~$\mu$m CMOS process and has been validated using (2D) damage localization on an SHM test bed.
\end{abstract}

%
\IEEEpeerreviewmaketitle

\section{Introduction}
\begin{figure}[!tbh]
	\centering
	\includegraphics[width=0.85\columnwidth]{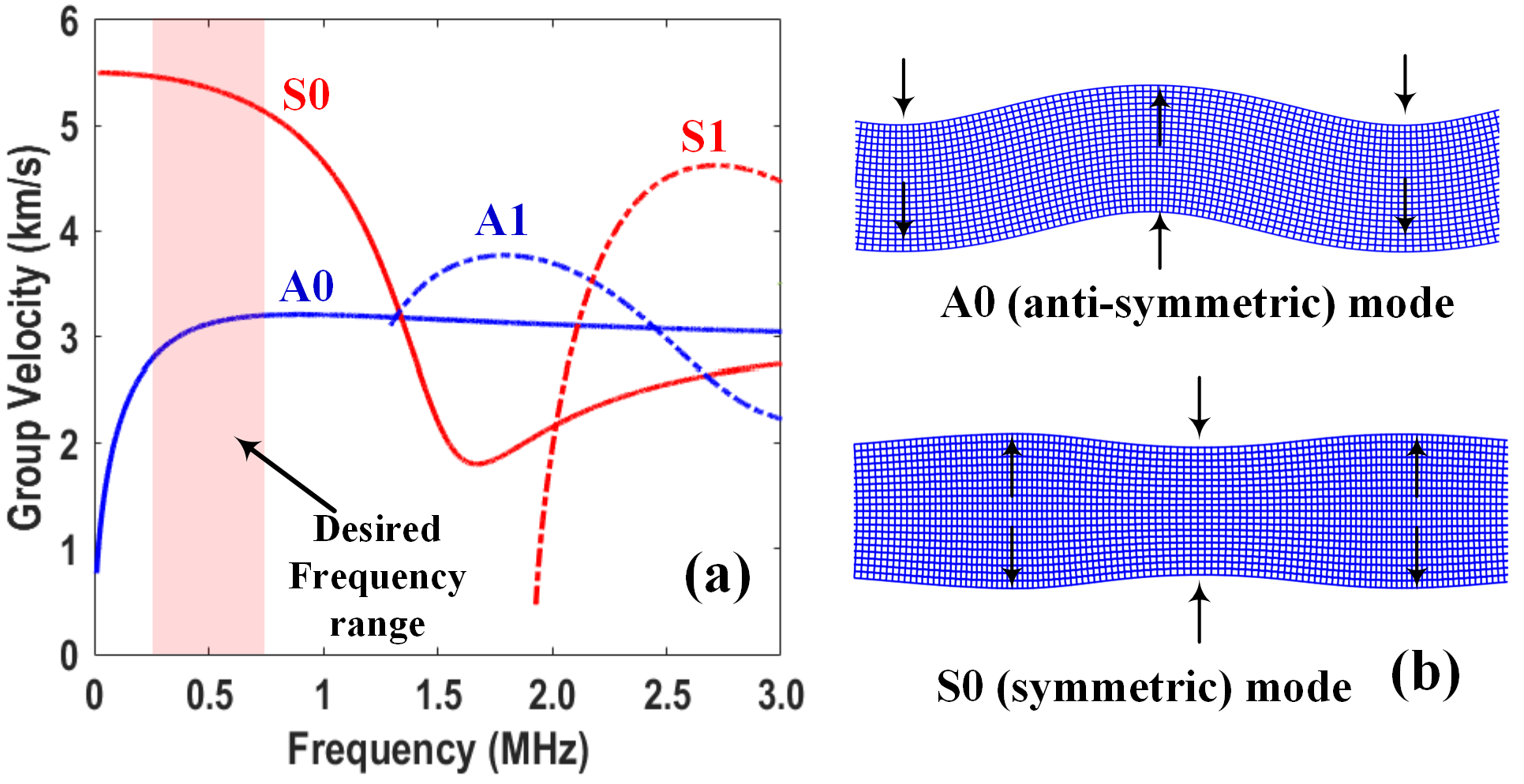}
	\caption{Dispersion of Lamb waves in a 1.5-mm-thick Aluminum plate: (a) variation in group velocity with frequency; (b) illustration of plate deformation in the $A_{0}$ and $S_{0}$ modes.}
	\label{fig:1}
\end{figure}

The health of the world's critical infrastructure, such as bridges, aircraft, and pipelines, is a subject of increasing concern. SHM enables real-time and continuous assessment of structural health by detecting the existence, location, and severity of potential damage. In active SHM, electrical pulses drive an array of piezoelectric transducers attached to the structure, which convert them to ultrasonic guided waves. These waves are received by another set of transducers and analyzed to determine structural health \cite{Raghavan2007}. Lamb waves have attractive propagation properties and are thus the most widely used ultrasonic guided waves for active SHM. However, a variety of Lamb wave modes with different velocities can propagate in the structure (see Fig.~\ref{fig:1}). The excitation waveform is usually designed to only excite the lowest-order modes ($S_{0}$ and $A_{0}$) in order to reduce the complexity of SHM signal processing. 

SHM systems are typically realized using discrete components. However, the size, weight, and power consumption of this approach makes it unattractive for emerging applications such as monitoring of aerospace structures \cite{Staszewski2009}. This issue has been addressed using integrated piezoelectric drivers \cite{Guo2014,Pierco2015} and flexible sheets that combine integrated circuits (ICs) and thin-film transistors for passive strain sensing \cite{Hu2014,Hu2014a}. We propose a heterogeneous microsystem that integrates miniaturized electronics and sensors within a flexible substrate to further reduce system thickness, weight, and power, thus paving the way for scalable large-area SHM. We have earlier i) designed and tested a current-controlled transceiver IC in 0.5~$\mu$m 2P/3M CMOS for this application \cite{Zamani2016}, and ii) verified its ability to \emph{detect} structural damage \cite{Tang2016}. Here we describe an improved digitally-controlled SHM transceiver and verify its ability to both \emph{detect} and \emph{localize} damage using a SHM test bed. 

\section{Chip Design}
\begin{figure}[!tbh]
	\centering
	\includegraphics[width = 0.85\columnwidth]{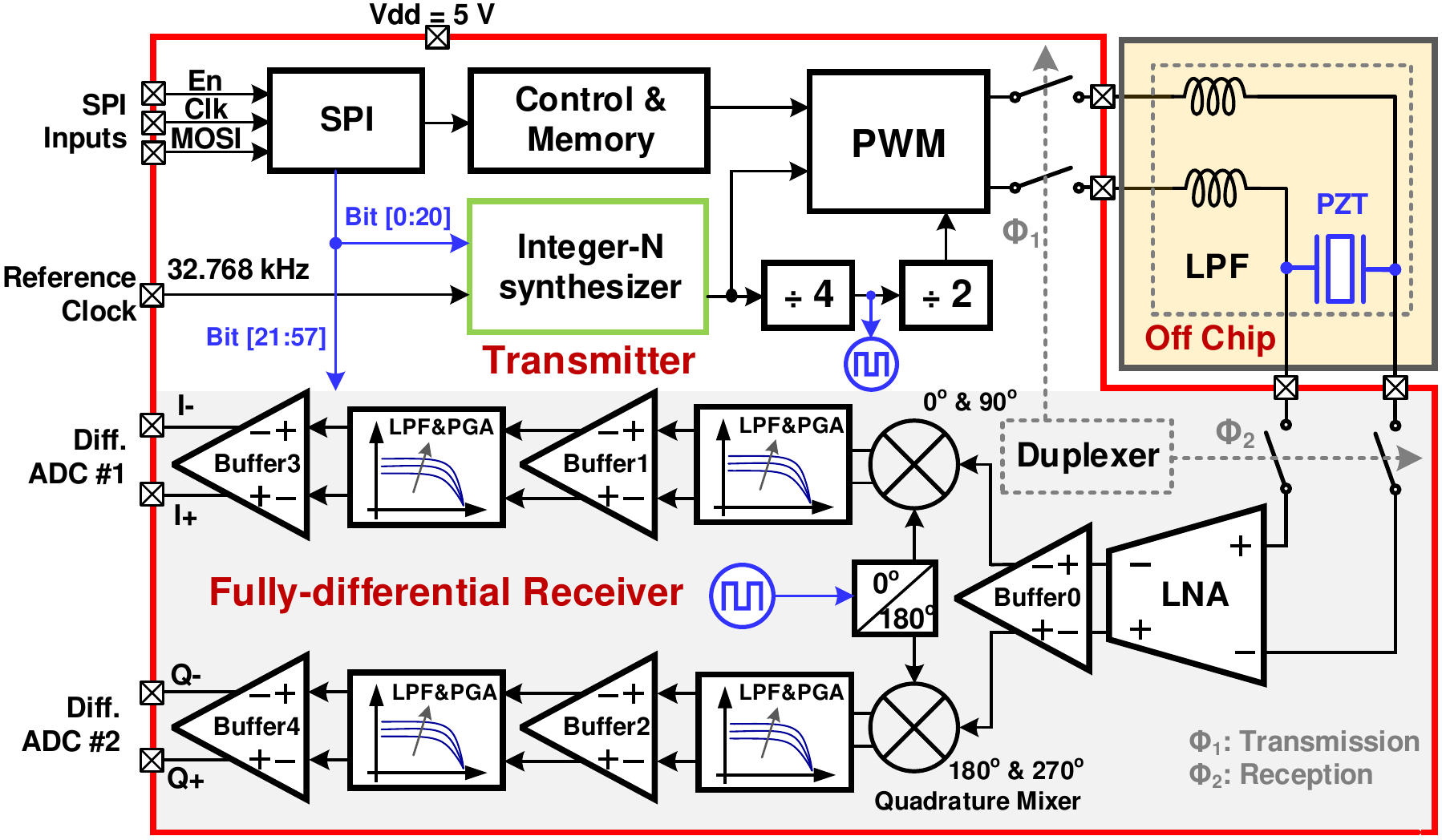}
	\caption{Block diagram of the proposed SHM transceiver IC.}
	\label{fig:2}
\end{figure}

A block diagram of the proposed IC is shown in Fig.~\ref{fig:2}. The transceiver runs off 5~V and is digitally-controlled through a standard serial peripheral interface (SPI) port. An integer-$N$ frequency synthesizer generates programmable output frequencies (0.8--22~MHz) from a  reference clock, thus allowing a single stable reference to be distributed to the entire SHM network. The synthesizer output is fed
into a 4-bit PWM modulator that generates predefined pulse widths based on data stored in on-chip memory (ROM). The widths are chosen to minimize the least-square error between the reference (a 5-cycle Hamming-windowed sinusoid) and the differential PWM outputs after a low-pass filter (LPF). The latter can drive signals up to $10(4/\pi)=12.7$~V$_{pp}$ into the transducer. 

The fully-differential receiver uses a low-noise amplifier (LNA) with programmable gain and bandwidth. The LNA outputs are down-converted to baseband using a passive double-balanced quadrature mixer. Two stages of low-pass filters and programmable-gain amplifiers are implemented for further filtering and amplification. The four complex outputs ($I^{+}$, $I^{-}$, $Q^{+}$, $Q^{-}$) retain both the amplitude and phase of the input signal, and are digitized by an off-chip ADC for further processing. The whole receiver is biased through a constant $G_{m}$ current reference and programmed over SPI. It can also be put into a low-power shutdown mode between pulses. 

\subsection{Low-Distortion Programmable Transmitter}
Modern active SHM algorithms, such as delay-and-sum and matched filtering, are implemented through baseline subtraction. Therefore, low-distortion and highly-accurate excitation waveforms are desired to reduce the false alarm rate (FAR) during long term SHM. Our solution uses (i) an integer-$N$ synthesizer based on a fourth-order loop filter that uses a low-phase-noise $ 32.768$~kHz reference clock; and (ii) a PWM scheme with predefined transitions based on \emph{a priori} knowledge of the desired SHM excitation signal after off-chip $LC$ low-pass filtering. This design allows the center frequency of the outputs to be programmed from 0.1-2.75~MHz, thus allowing a wide variety of structures to be studied. 

\begin{figure}[!tbh]
	\centering
	\includegraphics[width = 0.9\columnwidth]{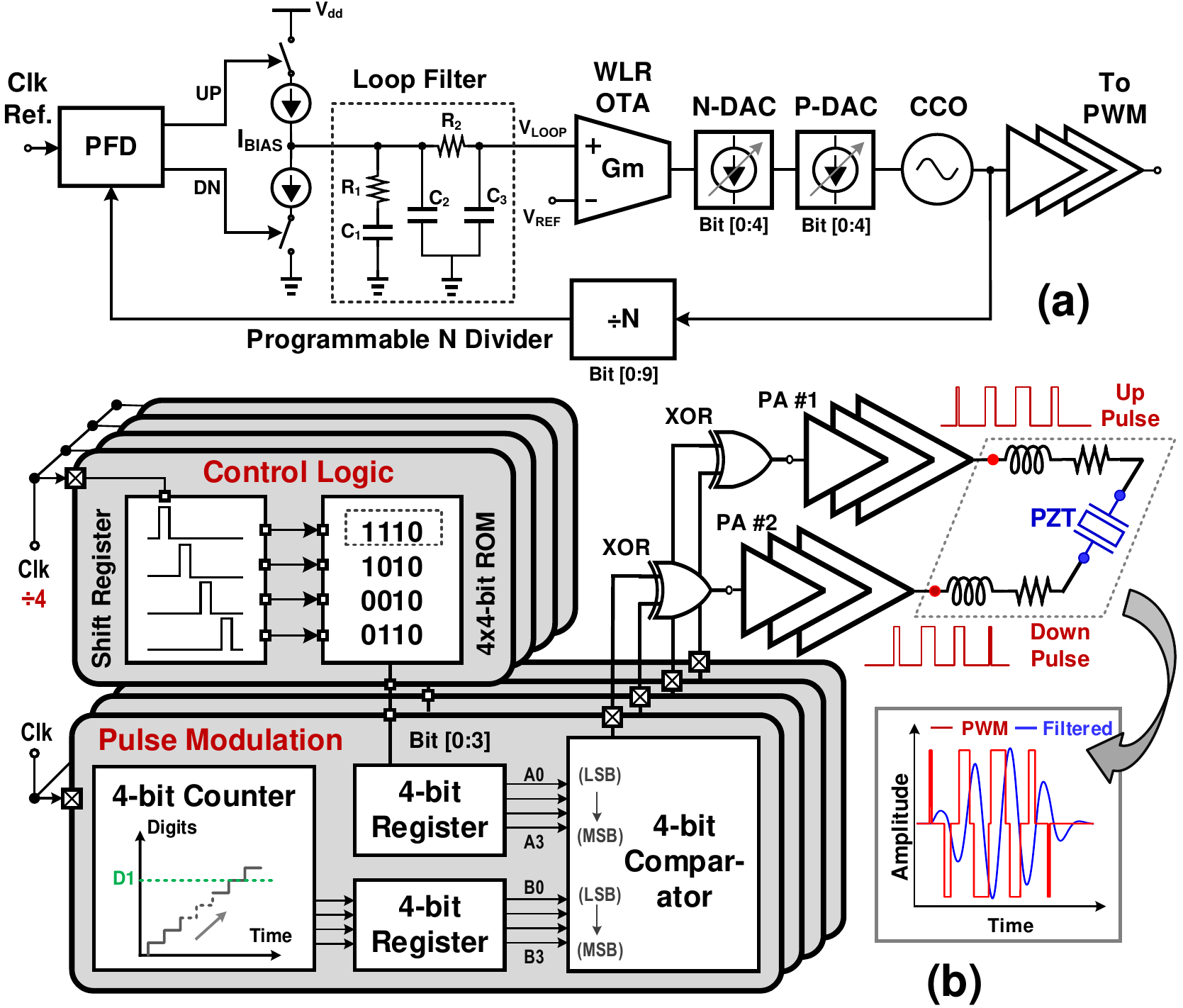}
	\caption{Low-distortion programmable transmitter circuit design: (a) block diagram of the integer-$N$ synthesizer; (b) digitally-defined PWM transition times.}
	\label{fig:3}
\end{figure}

Fig.~\ref{fig:3} shows the design of the main transmitter blocks in more detail. The synthesizer, shown in Fig.~\ref{fig:3}(a), includes a wide-linear-range transconductor (WLR), 5-bit NMOS and PMOS current DACs, and a current-starved ring oscillator (CCO). This allows the loop bandwidth to be controlled through the N- and P-DACs; in particular, we can make the bandwidth and phase margin independent of the division ratio ($N$). The synthesizer also uses a sequential phase frequency detector (PFD), a cascoded charge pump, and a passive third-order loop filter. The charge pump uses differential switching to reduce charge injection errors. The loop filter contains two high-frequency poles that are placed beyond the crossover frequency of the loop. These poles filter out high-frequency ripples on the control voltage ($V_{LOOP}$), thereby reducing jitter in the output clock while only minimally degrading phase margin. The WLR converts $ V_{LOOP}$ to a current; it combines a well-input differential pair with other linearization techniques to achieve $>1.5$~V input linear range. The loop bandwidth is set to 3.5~kHz (about 10\% of the nominal reference frequency) to avoid degradation of phase margin due to the phase lag inherent in a discrete-time PFD; the desired value is set to $ 50^{\circ}$. Fig.~\ref{fig:3}(b) shows the block diagram of the 4-bit PWM. To generate fully-differential excitation pulses, it uses two XOR gates and a $180^{\circ}$ delay circuit. The final pulse widths, as shown in Fig.~\ref{fig:3}(b), for the up-side and down-side are set as multiples of the synthesizer output period ($1$, $5$, $7$, $3$ and $3$, $7$, $5$, $1$, respectively). The PWM outputs are filtered using an off-chip $LC$ LPF in which the $C$ is provided by the transducer itself.

\begin{figure*}[t!]
	\centering
	\includegraphics[width = 0.85\textwidth]{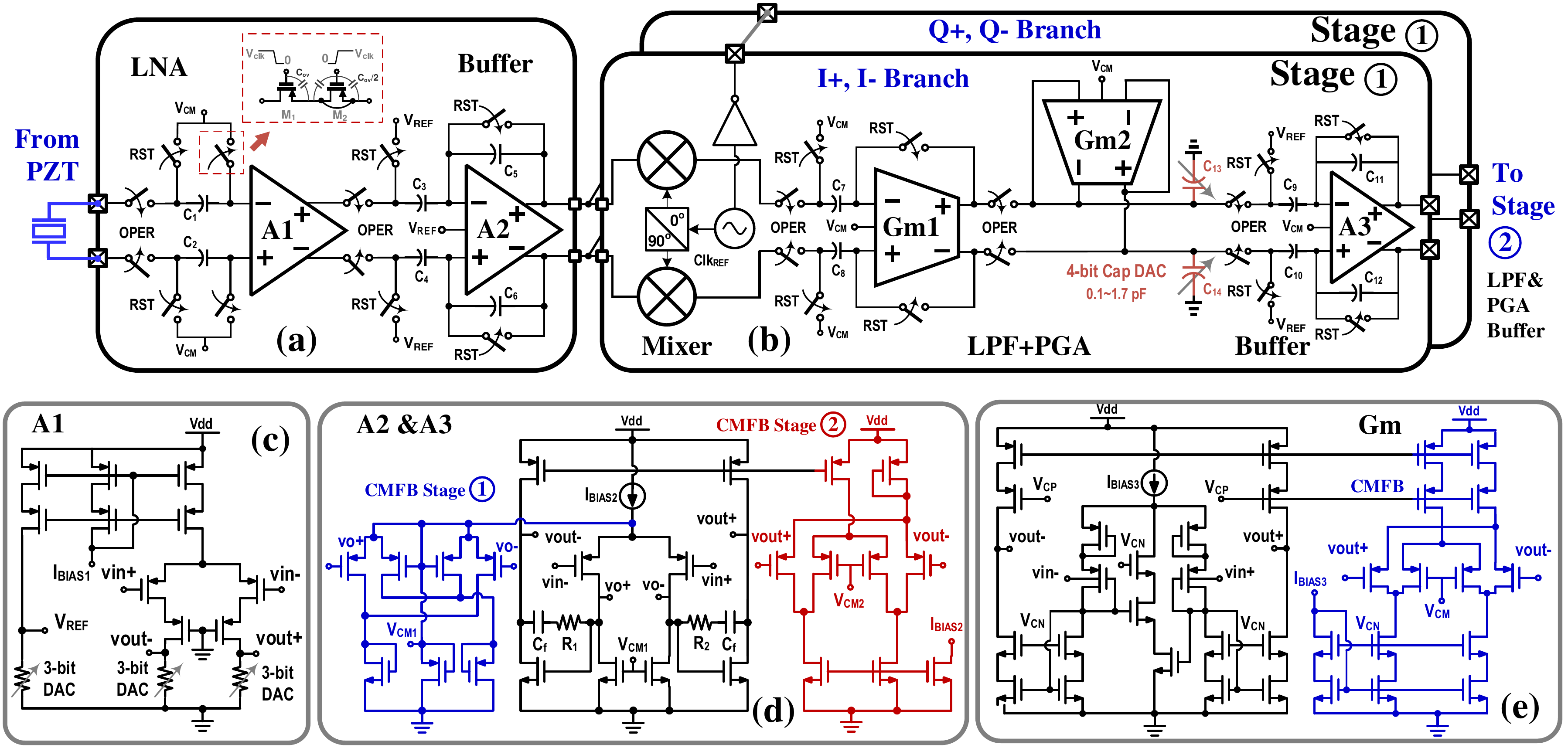}
	\caption{Block diagram of the fully-differential receiver and its circuit design: (a) LNA with low-leakage-current switches; (b) LPF and PGA blocks in stage $\#$1 after the passive mixers; (c) LNA circuit with programmable 3-bit resistor DAC; (d) fully-differential buffer with two stage CMFB; (e) fully-differential WLR OTA.}
	\label{fig:4}
\end{figure*}


\subsection{Offset-Canceling Fully-Differential Receiver}
The implementation of the SHM receiver is shown in Fig.~\ref{fig:4}. A trigger pulse $\Phi_{t}$ initiates the measurement. RST and OPER are two non-overlapping clock signals derived from $\Phi_{t}$. During the RST phase, each block resets to the input common-mode voltage $V _{CM}=1.3$~V. Auto-zeroing is used within each LPF and programmable-gain amplifier (PGA) for offset cancellation. The receiver operates during the OPER phase. 
  
A fully-differential cascoded low-noise amplifier (LNA), as shown in Fig.~\ref{fig:4} (a), is used as the first signal processing stage. The circuit uses a simple resistively-loaded differential pair and has an input-referred thermal noise PSD of $\overline{v_{n,in}^{2}}\approx 8kT\gamma/g_{m}$, where $\gamma\approx 2/3$ and $g_{m}$ is the transconductance of each transistor in the input pair. A replica bias circuit generates a reference voltage $V_{ref}$ that is nominally equal to the DC output voltage of the LNA. This voltage is used to set the common-mode input level of a differential buffer. We also use 3-bit resistor and bias current DACs for programmable LNA gain and bandwidth. Typically, at a bias current of 3~$\mu$A and a 40~k$\Omega$ load, the LNA has a simulated small-signal gain of 21.6~dB, a bandwidth of 4.3~MHz, a $1/f$ corner frequency of 10~kHz, and an input-referred thermal noise PSD of 21~nV/Hz$^{1/2}$. The input linear range, defined using total harmonic distortion (THD) $<5$\%, is $\sim$60 mV and $\sim$70 mV for input frequencies of 400~kHz and 1.2~MHz, respectively.
 
The differential buffer isolates LNA outputs from the passive double-balanced quadrature downconversion mixer (see Fig.~\ref{fig:4}(b) and (d)). The buffer uses two common-mode feedback (CMFB) amplifiers, one for each gain stage. A passive mixer design was chosen because of its small area, absence of static power consumption, and high linearity, while the double-balanced topology ensures high isolation between the ports. The typical conversion loss is $-4$~dB. Quadrature local oscillator (LO) signals for the mixers are generated from the same synthesizer used in the transmitter, resulting in a direct conversion (zero-IF) architecture. An external LO input also allows the receiver to be characterized at non-zero IF. 

A second-order $G_{m}$-$C$ LPF removes the upper mixer sidebands. It uses a fully-differential WLR OTA with continuous-time CMFB, as shown in Fig.~\ref{fig:4}(e). The LPF cut-off frequency can be programmed using a 5-bit current DAC and a 4-bit capacitor DAC (up to 1.5~pF). The PGA circuit reuses the same OTAs as in the LPF, with one acting as a $V$-$I$ converter and the other as a buffered resistor. The voltage gain $G=G_{m1}/G_{m2}\approx I_{Bias1}/I_{Bias2}$ is set by two 5-bit current DACs, where the approximation is valid in subthreshold.c

\begin{figure}[tbh!]
	\centering
	\includegraphics[width=0.75\columnwidth]{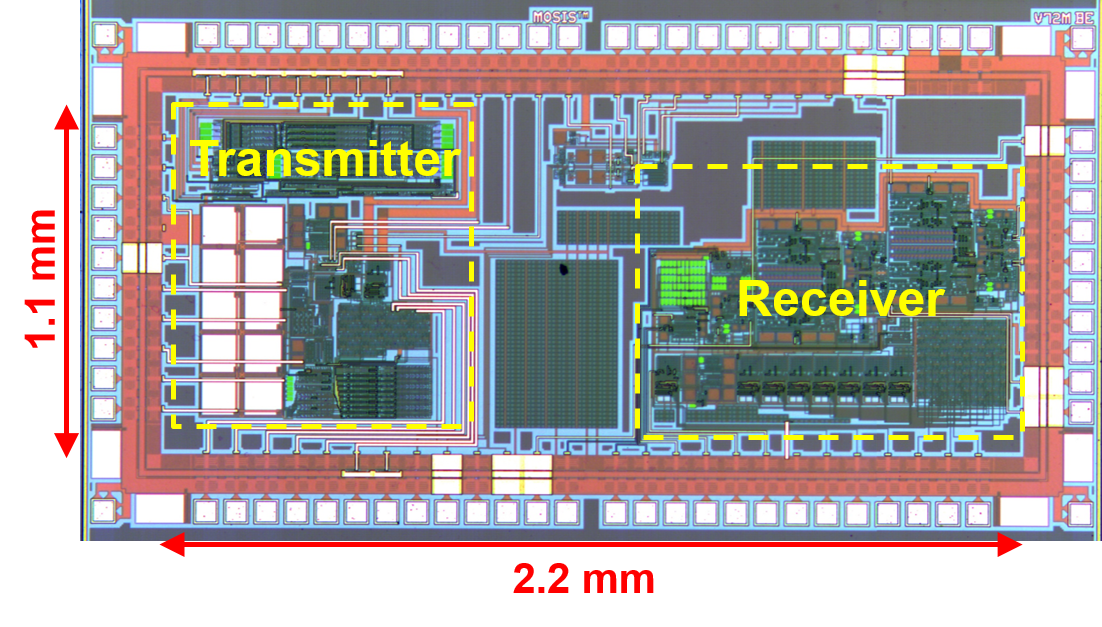}
	\caption{Die photograph of the proposed transceiver IC.}
	\label{fig:die_photo}
\end{figure}

\begin{figure}[tbh!]
	\centering
	\includegraphics[width=0.95\columnwidth]{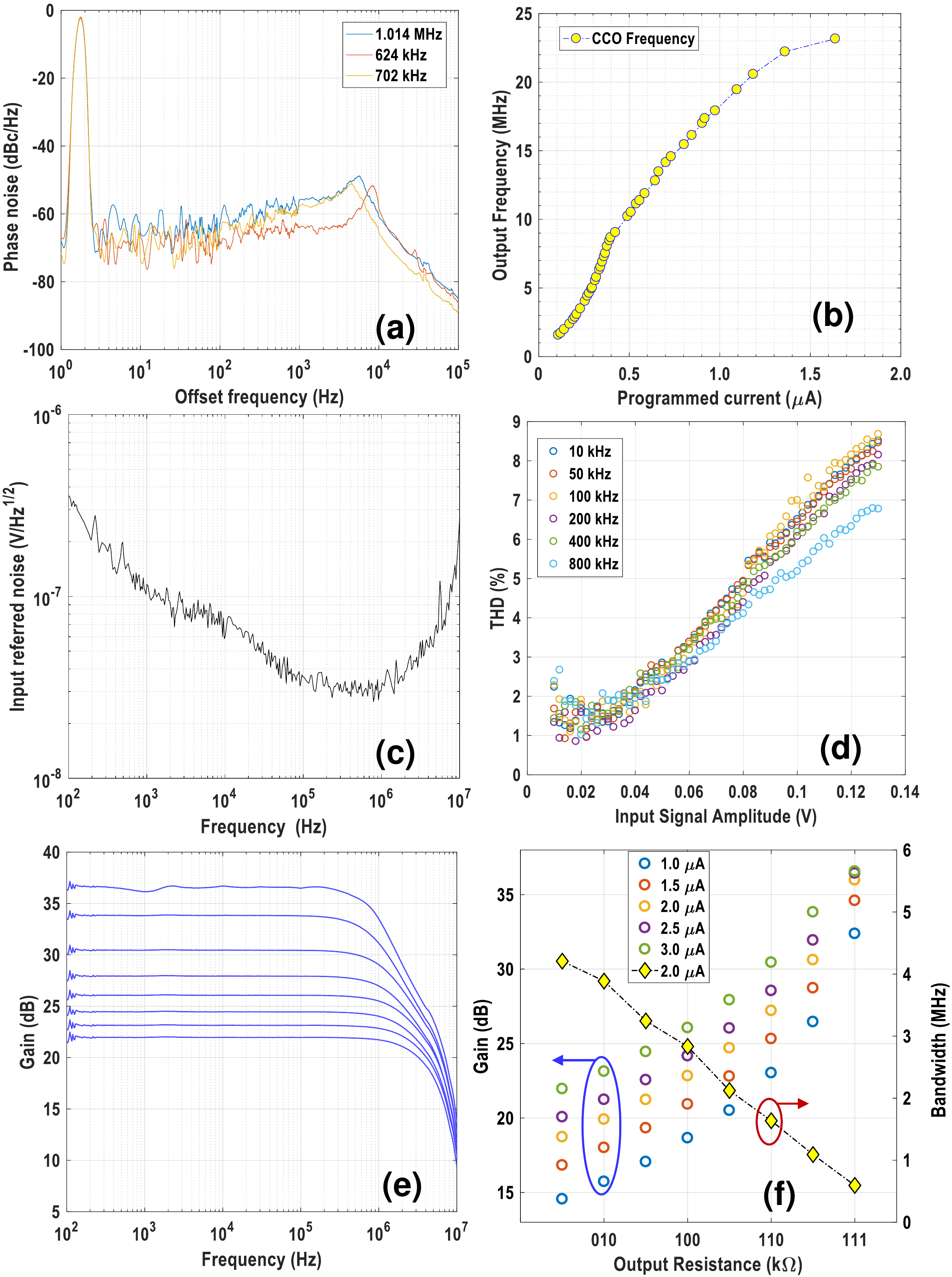}
	\caption{Chip measurement results: (a) phase noise for synthesizer output frequencies of 624~kHz, 702~kHz and 1.014~MHz; (b) CCO output frequency range; (c) input-referred noise of the LNA and buffer; (d) THD of the LNA and buffer; (e) measured gain of the LNA at 3.0~$\mu$A; (f) measured LNA gain and bandwidth in different scenarios.}
	\label{fig:6}
\end{figure}

\begin{figure}[tbh!]
	\centering
	\includegraphics[width = 0.95\columnwidth]{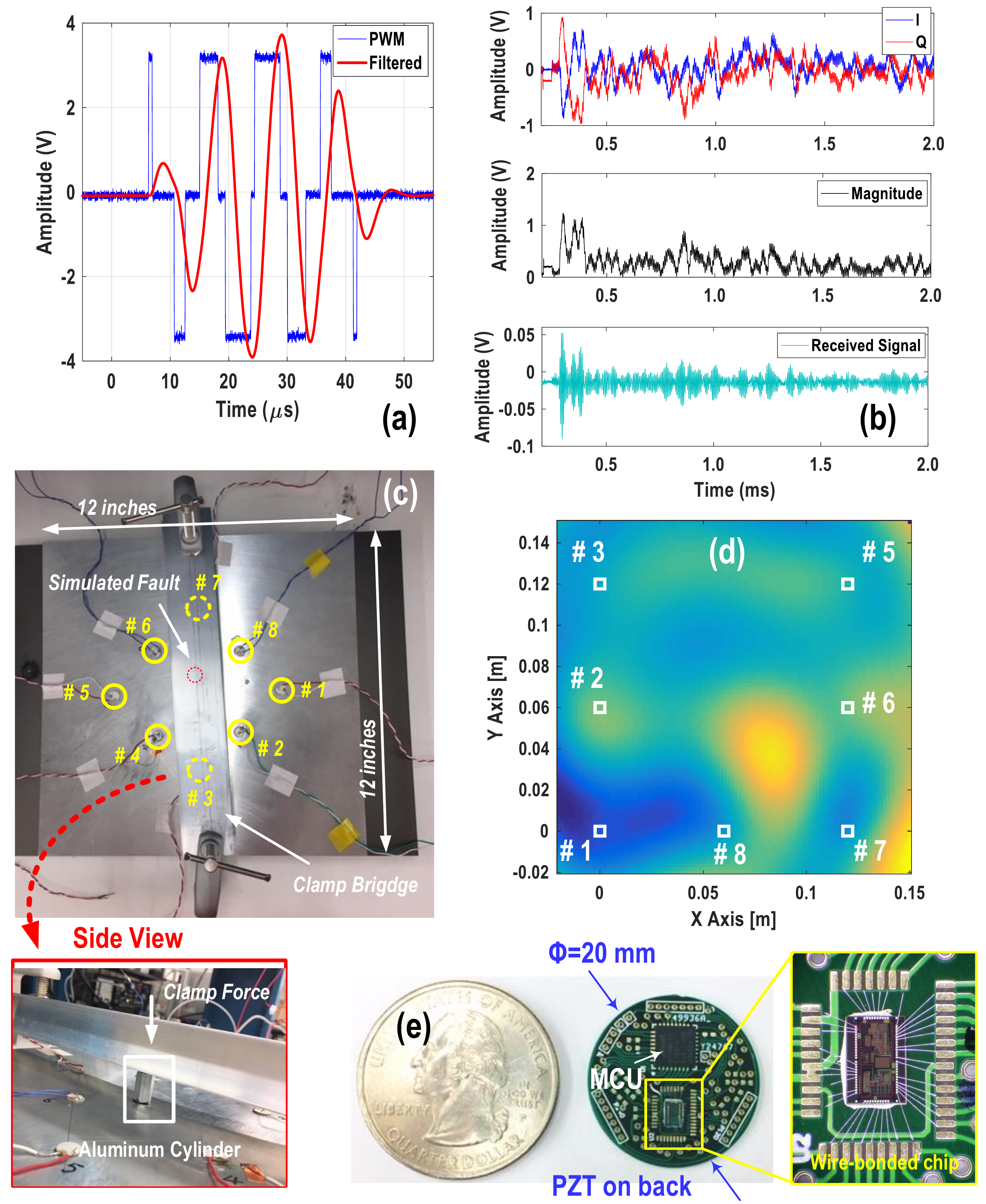}
	\caption{Experimental evaluation in a SHM test bed: (a) measured PWM digital output of the transmitter, and smooth Hamming-like waveform generated across the piezoelectric transducer after low-pass filtering; (b) typical $I$ and $Q$ outputs from the SHM receiver, reconstructed signal magnitude, and the raw signal; (c) experimental setup in an SHM test bed;(d) 2D damage localization results from the sensor node using delay-and-sum algorithm; (e) miniaturized SHM sensor node containing a commercial MCU and the SHM IC.}
	\label{fig:7}
\end{figure}

\section{Experimental Results}
A proof-of-concept SHM transceiver IC was fabricated in the OnSemi 0.5 $\mu$m CMOS process. Fig.~\ref{fig:die_photo} shows a die micrograph of the proposed IC, which has a active area of 1.1~mm $\times$ 2.2~mm. The chip is powered from a single 5.0~V supply with average power consumption of 875~$ \mu$W. Fig.~\ref{fig:6}(a) shows measured phase noise, which decreases to about -70~dBc/Hz at an offset of 10~Hz, while the transmitter tuning curve is shown in Fig.~\ref{fig:6}(b). Figs.~\ref{fig:6}(c) and (d) show the measured input-referred noise PSD and THD, respectively, of the LNA and buffer. The thermal noise floor and maximum input signal amplitude (defined by THD $<5$) are in good agreement with simulations. Given that the received signal amplitude in active SHM is typically $\sim$0.5~mV to 20~mV (depending on the sensor location), the proposed LNA provides sufficient linear range. 

\begin{figure}[tbh!]
	\centering
	\includegraphics[width = 0.75\columnwidth]{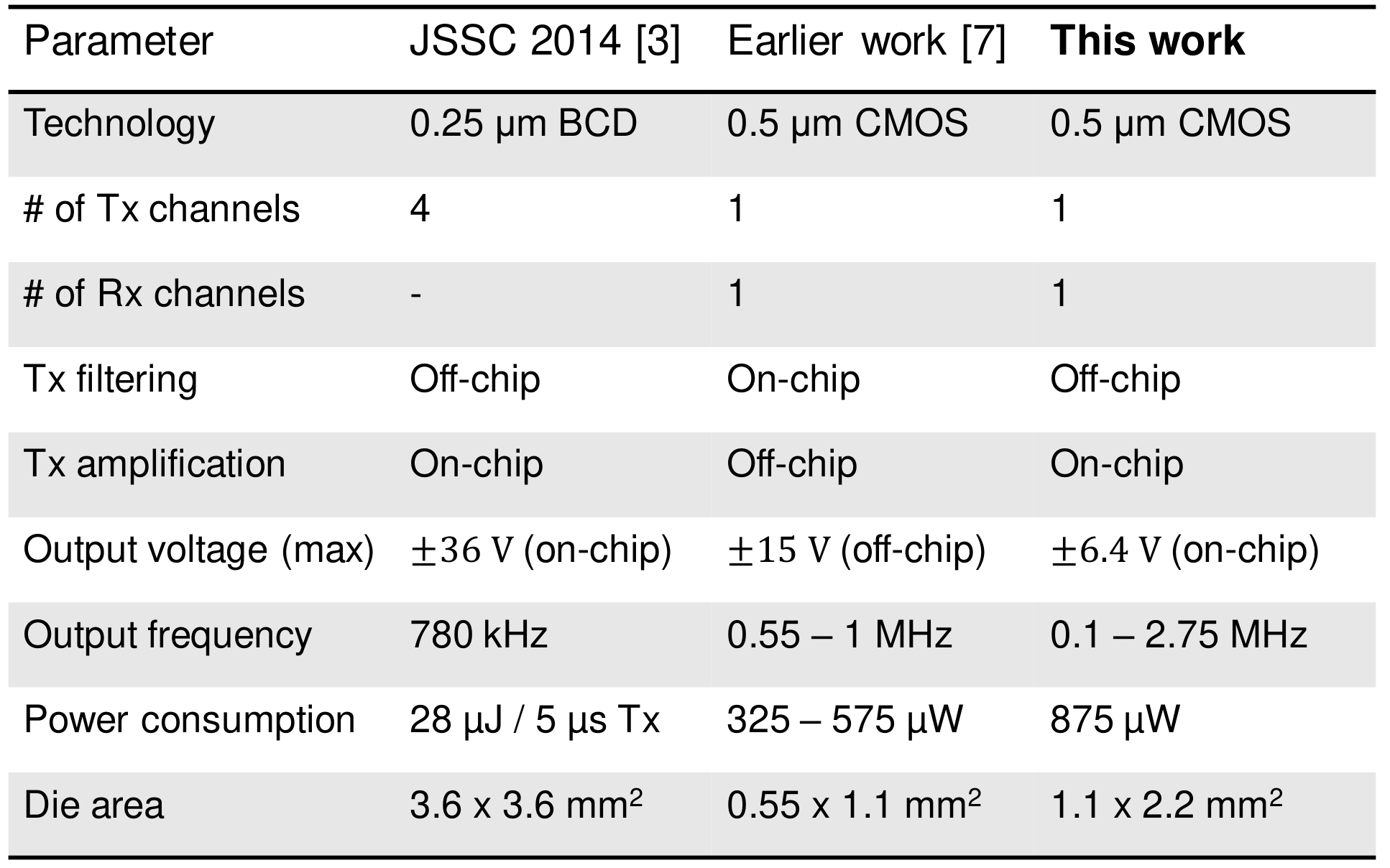}
	\caption{Comparison of the proposed chip with recently-reported driver ICs for SHM applications.}
	\label{fig:8}
\end{figure}

Fig.~\ref{fig:7} shows the IC connected to an SHM test bed (1 foot$ ^{2} $ area). The bare dies were wire-bonded to miniature (diameter~=~20~mm) wired sensor nodes fabricated (see Fig.~\ref{fig:7}(e)). Each sensor node also contains a microcontroller (MCU; Atemel SAM L21E) for programming, digitization, signal processing and communicating with the system controller. The PZT transducer ($6\times 6\times 0.5$~mm) is connected to the back of the PCB. Fig.~\ref{fig:7}(b) shows typical receiver outputs ($I$ and $Q$). The reconstructed signal amplitude follows the envelope of the input signals, as expected. Fig.~\ref{fig:7}(d) shows an example of successful 2D localization of structural damage using the custom IC and an array of 8 PZT transducers mounted on a 3~mm-thick aluminum plate. Reversible damage is introduced as an stressed point loaded by clamps (see Fig.~\ref{fig:7}(c)). The damage map was generated by using the delay-and-sum localization algorithm on data recorded from the IC. Fig.~\ref{fig:8} compares the performance of our chip with recently-reported driver ICs for active SHM applications \cite{Guo2014,Zamani2016}.

\section{Conclusion}
We have i) presented a digitally-programmable single-chip transceiver for active SHM using ultrasonic guided waves, and ii) successfully localized damage on a SHM test bed using the proposed chip and a delay-and-sum algorithm. Future work will focus on i) further miniaturization of the sensor nodes by integrating an analog-to-digital converter (ADC) and custom digital signal processor (DSP) on the chip; and ii) eliminating the wired bus by using ultrasonic power and data transfer. 
\bibliographystyle{IEEEtran}
\bibliography{IEEEabrv,CICC_2018}
%
%

\end{document}